\documentclass[fp,twocolumn]{jpsj3}
\usepackage[dvipdfmx]{graphicx}

\makeatletter 
\newcommand\newblock{\hskip .11em\@plus.33em\@minus.07em}
\makeatother

\usepackage{amsmath, amssymb}
\usepackage{bm}
\usepackage{mathrsfs}
\usepackage{physics}
\usepackage{comment}
\usepackage[normalem]{ulem}

\usepackage{color}

\title{Analysis of Spin Current Generation by Elastic Waves \\in $f$-wave Altermagnets}
\author{Ken Uchino$^*$, Yuuki Ogawa$^\dagger$, and Satoru Hayami$^\ddagger$}
\inst{Graduate School of Science, Hokkaido University, Sapporo 060-0810, Japan}

\abst{
We theoretically investigate the mechanism of spin current generation induced by elastic waves in nonrelativistic magnets referred to as altermagnets. 
By analyzing an $f$-wave altermagnet formed by a three-sublattice noncollinear antiferromagnetic structure breaking the spatial inversion symmetry on a two-dimensional triangular lattice within the linear response theory,
we show that the nonrelativistic antisymmetric spin-split band structure can give rise to spin current generation when either a longitudinal or a transverse elastic wave is applied.
We find that the momentum dependence of the antisymmetric spin splitting leads to a characteristic direction-dependent spin current response.
We also compare the present nonrelativistic magnetic-order-driven mechanism with the relativistic one in a nonmagnetic Rashba system. 
These findings highlight the potential of inversion-symmetry-breaking altermagnets as a spin current generator driven by elasticity without relying on the relativistic spin--orbit coupling.
}

\date{\today}
\begin{document}
\maketitle

\section{Introduction}
\label{intro}
In recent years, it has been discovered that certain antiferromagnetic materials with zero net magnetization can exhibit momentum-dependent spin-split band structures even in the absence of the relativistic spin--orbit coupling (SOC)~\cite{Noda2016, Okugawa2018, Ahn2019, Naka2019, Hayami_YK2019, Hayami_PhysRevB.102.144441, Yuan2020, Naka_PhysRevB.103.125114, Yuan2021PRMat, Yuan2021PRB}.
Such antiferromagnetic materials are now widely recognized as altermagnets~\cite{Gonzalez-Hernandez2021, Smejkal2022PRX1, Smejkal2022PRX3, Smejkal2022PRX4}, which have 
attracted considerable interest as a new class of spintronic materials that do not rely on strong SOC inherent in heavy elements.
~\cite{Kato2004, Ahn2019, Naka2019, Sun_PhysRevB.108.L140408, Bai_PhysRevLett.130.216701, Giil_PhysRevB.110.L140506, Ogawa2026}.
Not being restricted to the presence of the SOC, several intriguing phenomena free from the SOC have been suggested, such as the spin current generation~\cite{Ahn2019, Naka2019, hayami2022spinconductivity, Sourounis_PhysRevB.111.134448, Ezawa_PhysRevB.111.125420, naka2025altermagnetic}, linear and nonlinear piezomagnetic effects~\cite{Ma2021, Zhu2024, Aoyama2024, McClarty2024_PhysRevLett.132.176702, Wu2024, Yershov2024, ogawa2025nonlinear, Naka2025nonrelativistic, Chen2025, Huyen2025anisotropic}, 
superconducting phenomena~\cite{Fukaya2025}, and nonlinear magnetoelectric effect~\cite{Oike_PhysRevB.110.184407, sato2025quantum, Oike2025thermodynamic}.
Although the concept of altermagnets is also used to indicate antiferromagnets exhibiting the anomalous Hall effect in collinear antiferromagnets by the aid of the SOC~\cite{Solovyev_PhysRevB.55.8060, Sivadas_PhysRevLett.117.267203, vsmejkal2020crystal,samanta2020crystal, Naka_PhysRevB.102.075112, Hayami_PhysRevB.103.L180407, Chen_PhysRevB.106.024421, Roig_PhysRevB.110.144412, ohgata2024intrinsic}, we focus on nonrelativistic spin-splitter effects by ignoring the SOC in the following discussion.

From the symmetry viewpoint, the breaking of time-reversal symmetry is necessary to cause the symmetric spin-split band structures, where the even-order wave vector is coupled to the spin. 
Meanwhile, the breaking of spatial inversion symmetry is not required. 
According to the angle dependence, altermagnets are classified by a $d$-wave altermagnet with $k_\alpha k_\beta \sigma$, a 
$g$-wave altermagnet with $k_\alpha k_\beta k_\gamma k_\eta \sigma$, and so on, where $k_\alpha$ and $\sigma$ stand for the $\alpha=x,y,z$ component of the wave number and spin, respectively. 
Furthermore, collinear magnetic structures are sufficient to induce the symmetric spin-split band structures described above. 
These conditions broaden the scope of candidate antiferromagnetic materials, which leads to the experimental findings of several altermagnets, such as MnF$_2$~\cite{Yuan2020, Smejkal2022PRX3, Smejkal2022PRX4, Hariki_PhysRevB.110.L100402, Morano_PhysRevLett.134.226702}, MnTe~\cite{Krempasky2024, Lee2024, Osumi2024}, and CrSb~\cite{Ding2024, Reimers2024, Zeng2024}.

In parallel with altermagnets causing the symmetric spin splitting, another type of altermagnets has been clarified by focusing on the nonrelativistic antisymmetric spin-split band structure, where the odd-order wave vector is coupled to the spin~\cite{Hayami2020}.
Such an antisymmetric spin splitting is caused in noncollinear antiferromagnetic structures breaking spatial inversion symmetry, such as the 120$^{\circ}$ structure in a triangle unit~\cite{Hayami2020, Hayami_PhysRevB.102.144441} and the spiral spin structure~\cite{Hayami_PhysRevB.105.024413, Brekke_PhysRevLett.133.236703}.
These antiferromagnets exhibiting the antisymmetric spin splitting are referred to as $p$-wave, $f$-wave, and $h$-wave altermagnets depending on the rotational symmetry in momentum space~\cite{hellenes2023p}. 
For example, the momentum dependence in $f$-wave altermagnets is characterized as \( k_y (k_y^2 - 3 k_x^2)  \sigma_z \) while preserving the threefold rotational symmetry, where the spin polarization direction is locked perpendicular to the spin-moment direction in real space. 
Such inversion-symmetry-breaking altermagnets have been extensively studied in the contexts of multipoles~\cite{Hayami2020, hayami2024unified}, quantum transport theory~\cite{kokkeler2025quantum}, impurity-induced Friedel oscillations~\cite{Sukhachov_PhysRevB.110.205114}, Andreev reflection~\cite{Soori_PhysRevB.111.165413}, flat-band zero-energy states~\cite{Nagae_PhysRevB.111.174519, Hodt_PhysRevB.111.205416}, and 
nonreciprocal spin current,\cite{Ezawa_PhysRevB.111.125420}. 
Recently, electrical switching of $p$-wave altermagnets has been experimentally demonstrated in NiI$_2$~\cite{song2025electrical}

In the present study, we investigate another physical phenomenon triggered by inversion-symmetry-breaking altermagnets.
Specifically, we demonstrate the spin current generation by elastic waves in $f$-wave altermagnets by analyzing the tight-binding model with a three-sublattice noncollinear antiferromagnetic structure on a triangular lattice
, where the spin current generation by the electric field has been shown~\cite{Zelezny2017}.
We show that the antisymmetric spin splitting in the form of  \( k_y(k_y^2 - 3 k_x^2) \sigma_z \) naturally leads to an effective coupling between spin current and strain, which becomes the origin of the spin current generation by elastic waves. 
We also present the characteristic spin current response against the direction of the input elastic waves. 
Furthermore, we compare the present results driven by the magnetic mean field with the previous results in nonmagnetic Rashba systems, where the relativistic SOC plays a significant role~\cite{Ogawa2023}. 
Our results provide another mechanism of the spin current generation based on inversion-symmetry-breaking altermagnets, which broadens the scope of candidate materials toward spintronic applications.

The rest of this paper is organized as follows. 
In Sec.~\ref{sec:model}, we formulate a tight-binding model on a two-dimensional triangular lattice incorporating a three-sublattice noncollinear antiferromagnetic order that characterizes the $f$-wave altermagnetic state.
In Sec.~\ref{sec:electronic_structure}, we analyze the resulting electronic band structure and demonstrate the emergence of antisymmetric spin splitting in the absence of the relativistic
SOC, as a consequence of the inversion symmetry-breaking magnetic order.
In Sec.~\ref{sec:strain}, we introduce the effect of dynamical strain induced by elastic waves through the modulation of hopping amplitudes, and derive the spin current response function within the linear response theory.
In Sec.~\ref{sec:results}, we present numerical results for the elastic-wave-induced spin current, focusing on its dependence on chemical potential, damping factor, and propagation angle of the elastic wave. 
We also evaluate the contribution of intraband and interband effects, and discuss the underlying mechanism from a symmetry perspective.
Finally, Sec.~\ref{sec:summary} summarizes the main findings of this study and discusses their implications for potential material realizations and future applications.  
A comparative analysis with the nonmagnetic Rashba systems is provided in the Appendix.

\section{Model}
\label{sec:model}
\begin{figure}[t]
    \centering
    \includegraphics[width=1.0\linewidth]{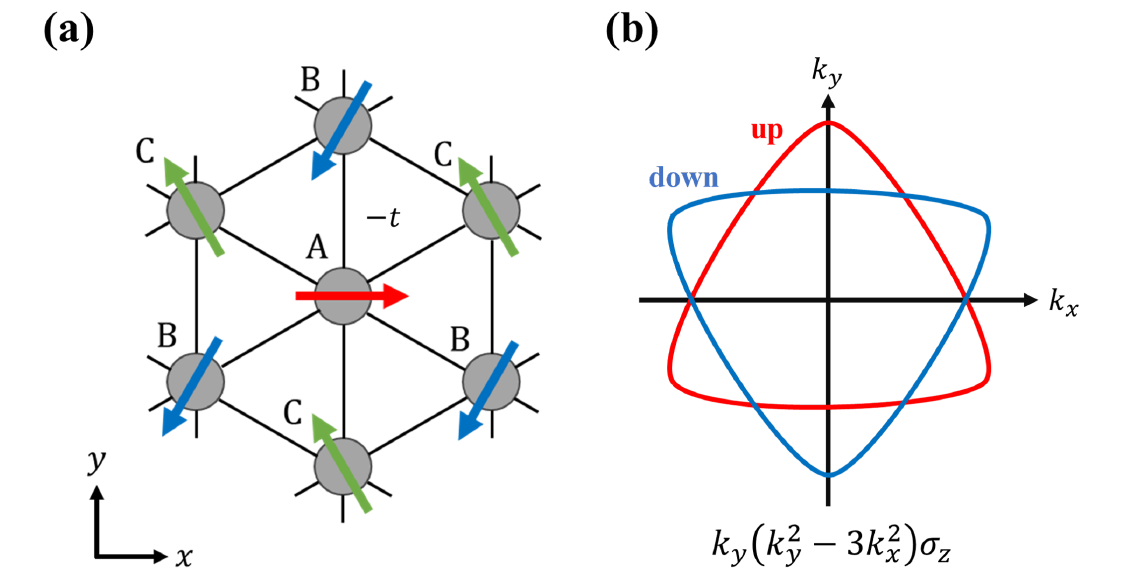}
    \caption{(Color online) (a) Three-sublattice noncollinear magnetic structures on the triangular lattice; three sublattices are denoted as A, B, and C. The arrows on each site represent the spin directions. $t$ stands for the nearest-neighbor hopping amplitude.
    (b) Schematic of the antisymmetric spin-split band structure in momentum space under the magnetic structures in (a); the momentum dependence is given by $k_y(k_y^2 - 3k_x^2)\sigma_z$.
    }
    \label{fig:model}
\end{figure}

We introduce a tight-binding Hamiltonian on a two-dimensional triangular lattice to examine the behavior of the spin current generation by elastic waves.
As a typical example, we consider a three-sublattice noncollinear antiferromagnetic ordering breaking spatial inversion symmetry, as shown in Fig.~\ref{fig:model}(a), which corresponds to $f$-wave altermagnets~\cite{Hayami2020, Hayami_PhysRevB.106.014420}.
Owing to the absence of the spatial inversion symmetry, the electronic band structure exhibits the antisymmetric spin splitting in momentum space of the form $k_y(k_y^2 - 3k_x^2)\sigma_z$
schematically shown in Fig.~\ref{fig:model}(b), as detailed below.

The model Hamiltonian is given by
\begin{equation}
H = H_{\mathrm{hop}} + H_{\mathrm{MF}},
\end{equation}
where the first term, $H_{\mathrm{hop}}$, represents the hopping integrals between the nearest-neighbor sites, whose expression reads
\begin{align}
H_{\mathrm{hop}} 
&= 
\sum_{\bm{k},\sigma}\sum_{\alpha,\beta = \mathrm{A}, \mathrm{B}, \mathrm{C}} \sum_{\bm{R}_{\alpha\beta}} 
E(\bm{R}_{\alpha\beta}) e^{i \bm{k} \cdot \bm{R}_{\alpha\beta}} 
c_{\bm{k}\alpha\sigma}^{\dagger} c_{\bm{k}\beta\sigma}\notag \\
&\equiv \sum_{\bm{k}}
c_{\bm{k}}^\dagger  \mathcal{H}_{\rm hop}(\bm{k}) c_{\bm{k}},
\label{eq:Hhop}
\end{align}

with the $6 \times 6$ matrix form
\begin{align}
\mathcal{H}_{\rm hop}(\bm{k})
=
\begin{pmatrix}
0 & \epsilon_{\bm{k}} \sigma_0 & \epsilon_{\bm{k}}^* \sigma_0\\
\epsilon_{\bm{k}}^* \sigma_0& 0 & \epsilon_{\bm{k}} \sigma_0\\
\epsilon_{\bm{k}} \sigma_0& \epsilon_{\bm{k}}^* \sigma_0& 0
\end{pmatrix};
\end{align}
\begin{align}
\epsilon_{\bm{k}}
= -t \left( e^{i \bm{k} \cdot \bm{\eta}_1}
+ e^{i \bm{k} \cdot \bm{\eta}_2}
+ e^{i \bm{k} \cdot \bm{\eta}_3}
\right).
\end{align}

Here, $c_{\bm{k}\alpha\sigma}^{\dagger}$ ($c_{\bm{k}\alpha\sigma}$) represents the creation (annihilation) operator of electrons with wave vector $\bm{k}$, sublattice $\alpha=$A--C, and spin $\sigma$, and $c_{\bm{k}}
={}^t( c_{\bm{k}\rm{A}\uparrow}$, $c_{\bm{k}\rm{A}\downarrow}$, $c_{\bm{k}\rm{B}\uparrow}$, $c_{\bm{k}\rm{B}\downarrow}$, $c_{\bm{k}\rm{C}\uparrow}$, $c_{\bm{k}\rm{C}\downarrow}$); $\sigma_0$ stands for the $2\times 2$ identity matrix in spin space.
$E(\bm{R}_{\alpha\beta})$ stands for the hopping amplitude between sublattices located at $\bm{r}_\alpha$ and $\bm{r}_\beta$, where $\bm{R}_{\alpha\beta} \equiv \bm{r}_{\beta} - \bm{r}_{\alpha}$ is the vector connecting their respective positions.
We suppose the nearest-neighbor hopping $-t \equiv E(\bm{R}_{\alpha\beta})$ for $|\bm{R}_{\alpha\beta}|=a$; $a$ is the lattice constant. 
The three nearest-neighbor vectors are denoted by $\bm{\eta}_1=a(\frac{\sqrt{3}}{2}\hat{x}-\frac{1}{2}\hat{y})$, $\bm{\eta}_2=a\hat{y}$, and $\bm{\eta}_3=a(-\frac{\sqrt{3}}{2}\hat{x}-\frac{1}{2}\hat{y})$, i.e., $\bm{R}_{\alpha\beta} =$ $\bm{\eta}_n$ or $-\bm{\eta}_n$ ($n = 1, 2, 3$).
We set $t=1$ and $a=1$ as the energy and length units, respectively.

The second term, $H_{\mathrm{MF}}=\sum_{\bm{k}} c_{\bm{k}}^\dagger \mathcal{H}_{\mathrm{MF}}  c_{\bm{k}}$, describes the antiferromagnetic mean field to induce the three-sublattice noncollinear magnetic structure, as shown in Fig.~\ref{fig:model}(a):
\begin{align}
\mathcal{H}_{\mathrm{MF}} 
= h
\begin{pmatrix}
\sigma_x & 0 & 0 \\
0 & -\frac{1}{2}\sigma_x - \frac{\sqrt{3}}{2}\sigma_y & 0 \\
0 & 0 & -\frac{1}{2}\sigma_x + \frac{\sqrt{3}}{2}\sigma_y 
\end{pmatrix},
\end{align}
where $\sigma_x$ and $\sigma_y$ are the Pauli matrices in spin space.
We here suppose that all the spin moments lie on the $xy$ plane without loss of generality. 
In the following analysis, we set the magnitude of the mean field to $h=1$.

\section{Electronic Structure}
\label{sec:electronic_structure}
\begin{figure}[t]
    \centering
    \includegraphics[width=1.0\linewidth]{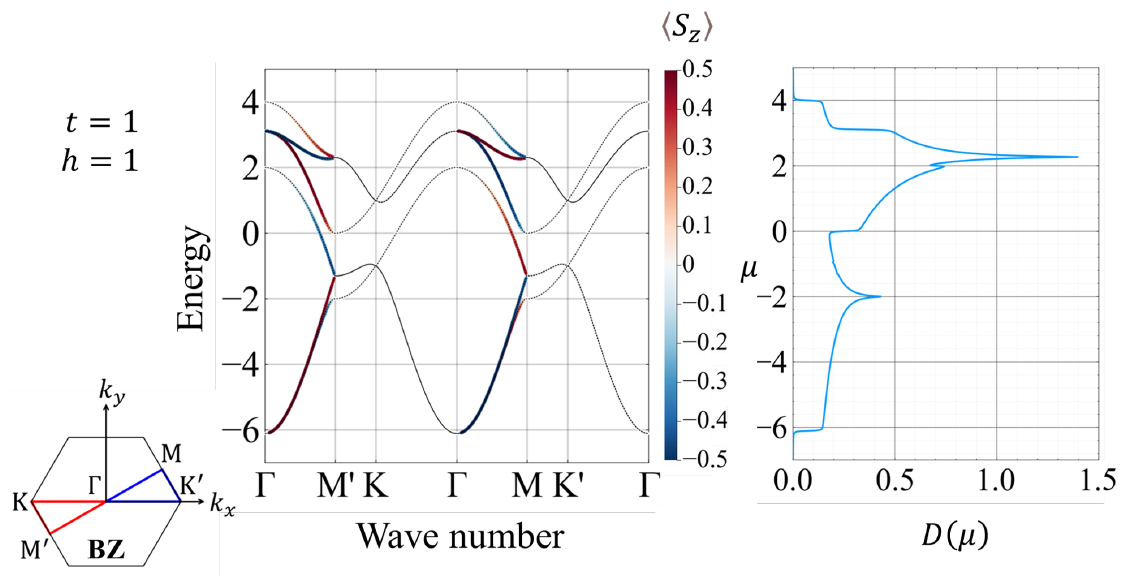}
    \caption{(Color online) (Left panel) Antisymmetric spin-split band structure along the high-symmetry lines in the Brillouin zone (BZ) under the three-sublattice noncollinear spin configuration.
    The color represents the expectation value of the momentum-resolved spin polarization along the $z$-axis. 
    (Right panel) Corresponding density of states.}
    \label{fig:band}
\end{figure}

The three-sublattice noncollinear antiferromagnetic ordering breaks the spatial inversion symmetry. 
Accordingly, such an ordered state brings about the antisymmetric spin-split band structure, as schematically shown in Fig.~\ref{fig:model}(b)~\cite{Hayami2020}. 
The left panel of Fig.~\ref{fig:band} shows the band structure for $h=1$, where the color represents the expectation value of the momentum-resolved spin polarization along the \(z\)-axis. 
The results show the appearance of the antisymmetric spin splitting along the high-symmetry lines like \(\Gamma\)--\({\rm M}\) and \(\Gamma\)--\({\rm M'}\)
, while it is absent along other directions like \(\Gamma\)--\({\rm K}\) and \(\Gamma\)--\({\rm K'}\). 
This behavior is consistent with an $f$-wave-type antisymmetric spin splitting of the form \(k_y(k_y^2 - 3k_x^2) \sigma_z\) depicted in Fig.~\ref{fig:model}(b).
It is noted that the present antisymmetric spin splitting is caused by noncollinear antiferromagnetic ordering instead of relativistic SOC.
This is in constrast to the nonmagnetic and collinear antiferromagnetic systems, where the SOC plays a significant role~\cite{li2013coupling, Hayami_PhysRevB.90.081115}.

We also show the corresponding density of states in the right panel of Fig.~\ref{fig:band}, where $\mu$ represents the chemical potential.
The peaks appear near the chemical potentials \(\mu = -2\) and \(\mu = 2.3\), which enhances the spin current induced by elastic waves, as described in Sec.~\ref{sec:results}.

\section{Calculation of Spin Current Induced by Elastic Waves}
\label{sec:strain}
We consider the effect of dynamical strain by elastic waves by the modulation of hopping integrals.
Following Ref.~\citen{Ogawa2023}, we adopt two assumptions for the modulation of the hopping amplitudes: One is that the strain-induced hopping modulation depends only on the interatomic distance, and is independent of direction. 
The other is that the hopping amplitude behaves inversely proportional to the square of the 
nearest-neighbor distance, i.e., $E(\bm{R}) \propto R^{-2}$ according to Harrison's rule~\cite{Froyen1979, Harrison1980}. 
Under these two assumptions, the lowest-order contribution of the strain-induced hopping modulation is given by
\begin{align}
    &\quad\  E\big(\bm{R}_{\alpha\beta} + \delta\bm{R}(\bm{r}_{\beta}) - \delta\bm{R}(\bm{r}_{\alpha})\big) \notag \\
    &\simeq -t \left[ 1 - 2 \frac{\bm{R}_{\alpha\beta}}{|\bm{R}_{\alpha\beta}|} \cdot \frac{\delta\bm{R}(\bm{r}_{\beta}) - \delta\bm{R}(\bm{r}_{\alpha})}{|\bm{R}_{\alpha\beta}|} \right],
\end{align}
where $\delta \bm{R}(\bm{r}) = \delta \bm{R}_{\bm{q}} e^{i(\bm{q}\cdot\bm{r} -\omega t)}$ is local atomic displacement by elastic wave; $\bm{q}$ and $\omega$ denote its wave vector and frequency, respectively.
Consequently, the hopping Hamiltonian $H_{\mathrm{hop}}$ is modulated as
$\tilde{H}_{\mathrm{hop}} = H_{\mathrm{hop}} + \delta H $, which includes
\begin{align}
    \delta H 
    &= 
    2t \sum_{\bm{k}, \sigma} \sum_{\alpha, \beta = \mathrm{A}, \mathrm{B}, \mathrm{C}} \sum_{\bm{R}_{\alpha\beta}} e^{i \bm{k} \cdot \bm{R}_{\alpha\beta}} 
    \notag \\
    &
    \quad \times \left( l^2 u_{xx} + 2 l m u_{xy} + m^2 u_{yy} \right) e^{-i\omega t}
    c_{\bm{k}\alpha\sigma}^\dagger c_{\bm{k}\beta\sigma} \notag \\
    &\equiv \sum_{\bm{k}} 
    c_{\bm{k}}^\dagger
    \delta \mathcal{H}(\bm{k};\bm{q}) \,  c_{\bm{k}} 
    e^{-i\omega t}  ,
    \label{eq:modu_diff}
\end{align}
\begin{align}
\delta\mathcal{H}(\bm{k};\bm{q})
=
\begin{pmatrix}
0 & \delta\epsilon_{\bm{k};\bm{q}}\sigma_0 & \delta\epsilon_{\bm{k};\bm{q}}^*\sigma_0\\
\delta\epsilon_{\bm{k};\bm{q}}^*\sigma_0 & 0 & \delta\epsilon_{\bm{k};\bm{q}}\sigma_0 \\
\delta\epsilon_{\bm{k};\bm{q}}\sigma_0 & \delta\epsilon_{\bm{k};\bm{q}}^*\sigma_0 & 0
\end{pmatrix},
\end{align}
\begin{align}
\delta\epsilon_{\bm{k};\bm{q}}
&= 
t
\biggl[ \frac{3}{2}
\Bigl( 
e^{i \bm{k} \cdot \bm{\eta}_1} + e^{i \bm{k} \cdot \bm{\eta}_3}
\Bigr)u_{xx}
-\sqrt{3}
\Bigl( 
e^{i \bm{k} \cdot \bm{\eta}_1} - e^{i \bm{k} \cdot \bm{\eta}_3}
\Bigr)u_{xy}
\nonumber \\
& \quad\quad
+
\frac{1}{2}\Bigl( 
e^{i \bm{k} \cdot \bm{\eta}_1} + 4 e^{i \bm{k} \cdot \bm{\eta}_2} + e^{i \bm{k} \cdot \bm{\eta}_3}
\Bigr)u_{yy}
\biggr].
\end{align}
where $(l, m) = \bm{R}_{\alpha\beta}/ \abs{\bm{R}_{\alpha\beta}}$ denote the direction cosines of $\bm{R}_{\alpha\beta}$, satisfying $l^2 + m^2 = 1$, and  
$u_{ij}=\frac{1}{2}(\pdv{}{x^i}\delta R^j + \pdv{}{x^j}\delta R^i)_{\bm{q}} \simeq \frac{1}{2}(iq_i \delta R^j_{\bm{q}} + iq_j \delta R^i_{\bm{q}})$
represents the strain tensor.
In deriving Eq.~(\ref{eq:modu_diff}), we adopt a continuum description under the long-wavelength condition $qa \ll 1$, 
retaining only the $\bm{q}$-linear term included in the strain tensor $u_{ij}$. 
We suppose that the lattice is stable under this continuum approximation and neglect the higher-order terms, e.g., sixth order, that affect the stabilization of the triangular lattice.

We evaluate the spin current under the dynamical strain within the linear response theory~\cite{Kubo1957}.  
The corresponding complex susceptibility is given by  
\begin{align}
    \chi_{i}^{\zeta}(\bm{q}, \omega)
    &= \frac{1}{N} \sum_{\bm{k}} \sum_{m,n} 
    \frac{f(\varepsilon_{m\bm{k}}) - f(\varepsilon_{n\bm{k}})}{\omega + \varepsilon_{m\bm{k}} - \varepsilon_{n\bm{k}} + i\delta} \notag \\
    &\quad\qquad\qquad \times \bra{m\bm{k}} j_{i}^{\zeta} \ket{n\bm{k}} \bra{n\bm{k}} \delta 
    \mathcal{H}(\bm{k};\bm{q})
    \ket{m\bm{k}},
    \label{eq:spin_current_response_1}
\end{align}
where \( f(E) = [e^{
(E - \mu)/T}+ 1]^{-1} \) is the Fermi--Dirac distribution function, with the temperature set to \(T=10^{-3}\).  
$\varepsilon_{n\bm{k}}$ and $\ket{n\bm{k}}$ represent the eigenvalue and eigenstate 
of the unperturbed Hamiltonian 
$\mathcal{H}(\bm{k}) = \mathcal{H}_{\mathrm{hop}}(\bm{k}) + \mathcal{H}_{\mathrm{MF}}$ for band index $n$ and wave vector $\bm{k}$, respectively.
\(\delta\) represents the phenomenological broadening factor corresponding to the inverse of the relaxation time, $N$ represents the number of total sites, and we set $\hbar = 1$ and $k_{\rm B} =1$.
The spin current operator is defined as 
$j_i^{\zeta} (\bm{q}) = \sum_{\bm{k}} 
c_{\bm{k}-\bm{q}/2}^\dagger
\, j_i^{\zeta} \, 
c_{\bm{k}+\bm{q}/2}$ 
with
\begin{align}
    j_{i}^{\zeta}
    \equiv \frac{1}{2} 
    \left\{ \frac{\partial
    \mathcal{H}(\bm{k})}{\partial k_i}, 
    S_{\zeta} \right\}_+ ,
    \label{eq:spin_current_response_2}
\end{align}
\begin{align}
S_{\zeta}\equiv\frac{1}{2}\begin{pmatrix}
\sigma_{\zeta} & 0 & 0 \\
0 & \sigma_{\zeta} & 0 \\
0 & 0 & \sigma_{\zeta}
\end{pmatrix},
\end{align}
where \( \zeta \) denotes the spin polarization direction and \( i \) represents the propagation direction of the spin current.
It is noted that we ignore the $\pdv{\delta \mathcal{H}}{k_i}$ part, which contributes 0-th order of $\omega$ for the response, since we focus on the $\omega$-linear term 
of the spin current.

The $\omega$-linear term is obtained by expanding the susceptibility in terms of $\omega$ as follows:
\begin{align}
    \Phi_{i}^{\zeta}&\equiv \lim_{\omega\rightarrow 0} [\chi_{i}^{\zeta}(\bm{q}, \omega) - \chi_{i}^{\zeta}(\bm{q}, 0)]/\omega \notag \\
    &=
    -\frac{1}{N}\sum_{\bm{k}} \sum_{m,n} 
    \frac{f(\varepsilon_{m\bm{k}}) - f(\varepsilon_{n\bm{k}})}{\varepsilon_{m\bm{k}} - \varepsilon_{n\bm{k}}} \notag \\
    &\quad \qquad
    \times \frac{\bra{m\bm{k}} j_{i}^{\zeta} \ket{n\bm{k}} \bra{n\bm{k}} \delta \mathcal{H}(\bm{k};\bm{q})\ket{m\bm{k}}}{\varepsilon_{m\bm{k}} - \varepsilon_{n\bm{k}} + i\delta}
\end{align}
where \(\Phi\) dominantly characterizes the efficiency of spin current generation by the applied elastic waves.  
The induced AC spin current is oscillating at the same frequency as the elastic wave. 
The AC spin currents can be experimentally detected through spin-to-charge conversion by the inverse spin Hall effect~\cite{Saitoh2006} 
or via magnon-mediated dynamic processes~\cite{Chumak2015}.
As a specific form of elastic waves, we consider a longitudinal wave defined as components
\begin{align}
    u_{xx} = u\cos^2\theta, \ 
    u_{yy} = u\sin^2\theta, \ 
    u_{xy} = u_{yx} = \frac{1}{2}u\sin2\theta,
\end{align}
where \( u \) denotes the strain-field magnitude, approximately equal to the ratio of the displacement amplitude to the lattice constant, 
and \( \theta \) is the propagation angle of the longitudinal wave measured from the \( x \)-axis.
We confirmed that the following results are not altered for a transverse wave at the qualitative level.

\section{Results}
\label{sec:results}
\begin{figure}[t]
    \centering
    \includegraphics[width=1.0\linewidth]{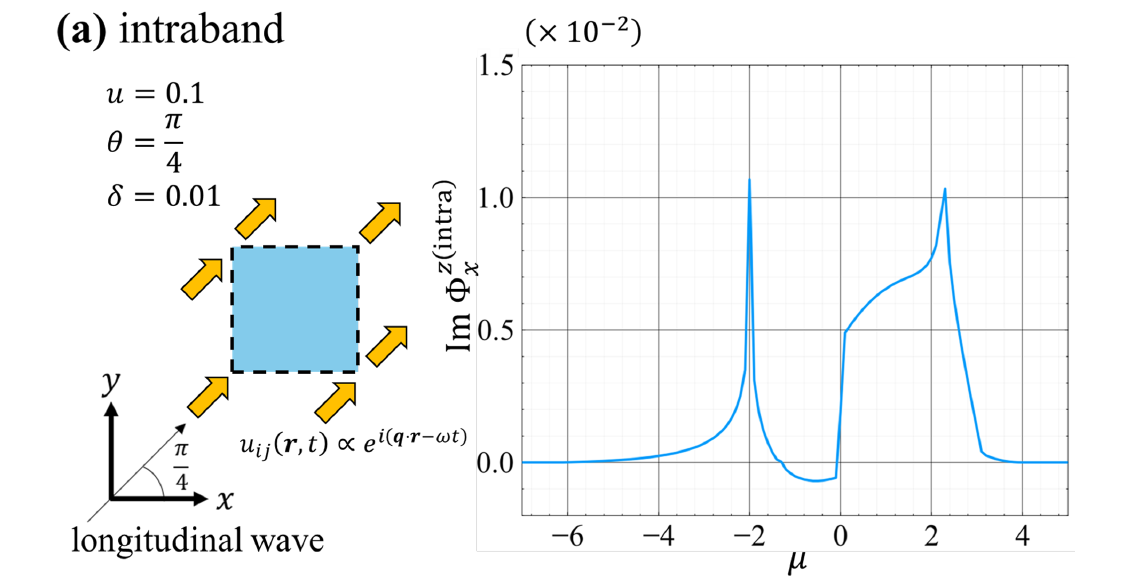} 
    \includegraphics[width=1.0\linewidth]{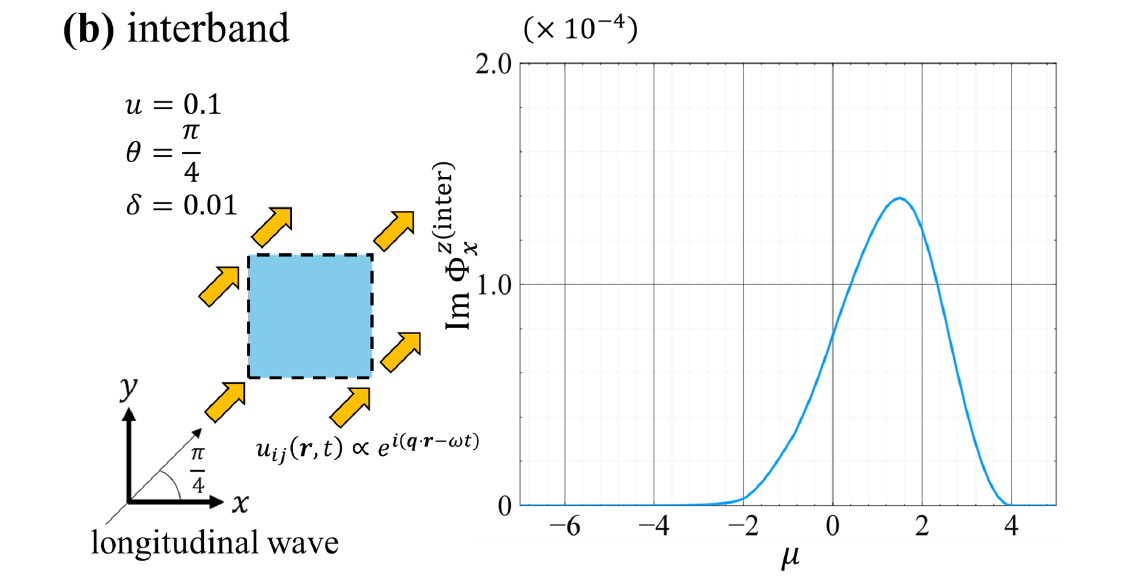} 
    \caption{(Color online) Chemical potential $\mu$ dependence of the spin current response by the application of a longitudinal elastic wave in the \(f\)-wave altermagnet, showing (a) intraband and (b) interband contributions.
    The left panel shows the propagation direction of the longitudinal wave.}
    \label{fig:mu}
\end{figure}

Figure~\ref{fig:mu} shows the chemical potential $\mu$ dependence of the spin current along the \(x\) 
direction with the $z$-spin polarization at $u=0.1$, $\theta=\pi/4$, and $\delta=0.01$; see the left panel of Figs.~\ref{fig:mu}(a) and \ref{fig:mu}(b) for the schematic setup.
To discuss the behavior of the spin current induced by elastic waves in detail, we decompose $\Phi^z_x$ into the intraband contribution $\Phi^{z{\rm (intra)}}_x$ with $\varepsilon_{n\bm{k}}=\varepsilon_{m\bm{k}}$ and the interband contribution $\Phi^{z{\rm (inter)}}_x$ with $\varepsilon_{n\bm{k}} \neq \varepsilon_{m\bm{k}}$. 
As the linear term in $\omega$ in the response function is purely imaginary, we analyze the imaginary part in the following.

Figure~\ref{fig:mu}(a) presents the results for ${\rm Im}[\Phi^{z{\rm (intra)}}_x]$ at $u=0.1$, $\theta=\pi/4$, and $\delta=0.01$. 
This contribution becomes nonzero for any $\mu$, which indicates that the spin current generation occurs under the application of a longitudinal wave in $f$-wave altermagnets.
In addition, one finds that ${\rm Im}[\Phi^{z{\rm (intra)}}_x]$ exhibits pronounced peaks near \(\mu = -2.0\) and \(\mu = 2.3\) 
, which is attributed to the large density of states, as shown in the right panel of Fig.~\ref{fig:band}.
Moreover, ${\rm Im}[\Phi^{z{\rm (intra)}}_x]$ tends to be larger when the Fermi level is located in a region with large antisymmetric spin splitting [left panel of Fig.~\ref{fig:band}], which is confirmed by comparing the regions for $-6 \lesssim \mu \lesssim -3$ and $0 \lesssim \mu \lesssim 4$; in the former region, 
antisymmetric spin splitting is almost absent, resulting in a small value of ${\rm Im}[\Phi^{z{\rm (intra)}}_x]$, whereas in the latter region, prominent antisymmetric spin splitting emerges, leading to a large value of ${\rm Im}[\Phi^{z{\rm (intra)}}_x]$. 
This is attributed to the contribution from only the intraband process in ${\rm Im}[\Phi^{z{\rm (intra)}}_x]$. 
Thus, the large antisymmetric spin splitting and the density of states at the Fermi level play an important role in enhancing the spin current response induced by elastic waves.

On the other hand, Fig.~\ref{fig:mu}(b) representing the interband contribution ${\rm Im}[\Phi^{z{\rm (inter)}}_x]$ shows a different tendency.
${\rm Im}[\Phi^{z{\rm (inter)}}_x]$ exhibits a broad peak structure around $\mu \simeq 1.5$, which is different from the peak positions for ${\rm Im}[\Phi^{z{\rm (intra)}}_x]$. 
This indicates that the density of states is not directly related to ${\rm Im}[\Phi^{z{\rm (inter)}}_x]$, which is captured by the fact that only the interband process contributes to ${\rm Im}[\Phi^{z{\rm (inter)}}_x]$. 
Meanwhile, the effect of the antisymmetric spin splitting plays an important role, similar to the case of ${\rm Im}[\Phi^{z{\rm (intra)}}_x]$; for example, ${\rm Im}[\Phi^{z{\rm (inter)}}_x]$ is extremely suppressed in the region for $-6 \lesssim \mu \lesssim -3$, where the antisymmetric spin splitting is invisible, as shown in the left panel of Fig.~\ref{fig:band}. 
The magnitude of the interband component is much smaller than that of the intraband component. 

\begin{figure}[t]
    \centering
    \includegraphics[width=1.0\linewidth]{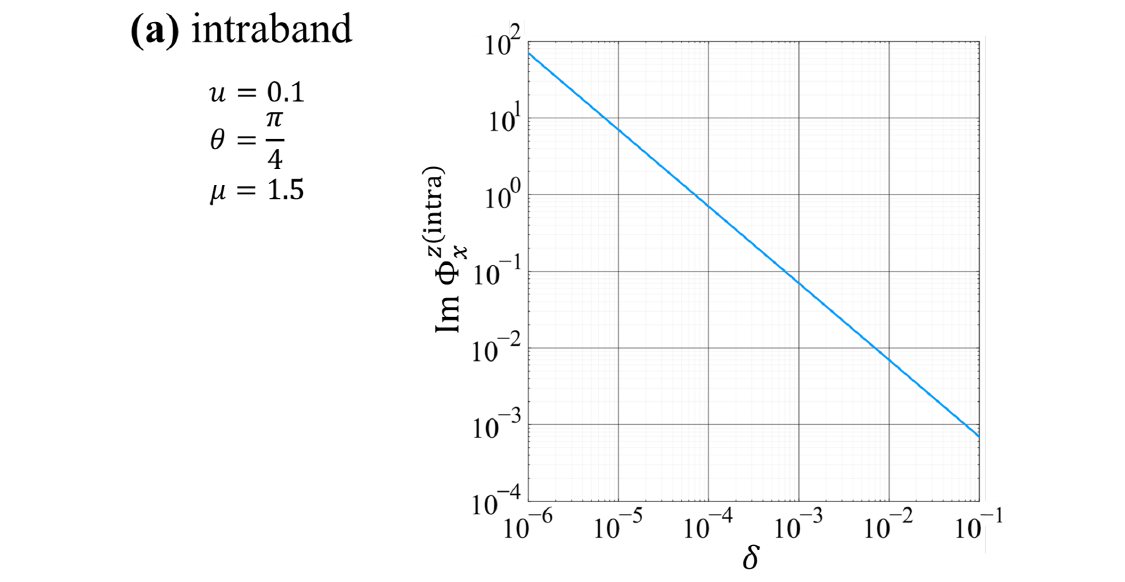} 
    \includegraphics[width=1.0\linewidth]{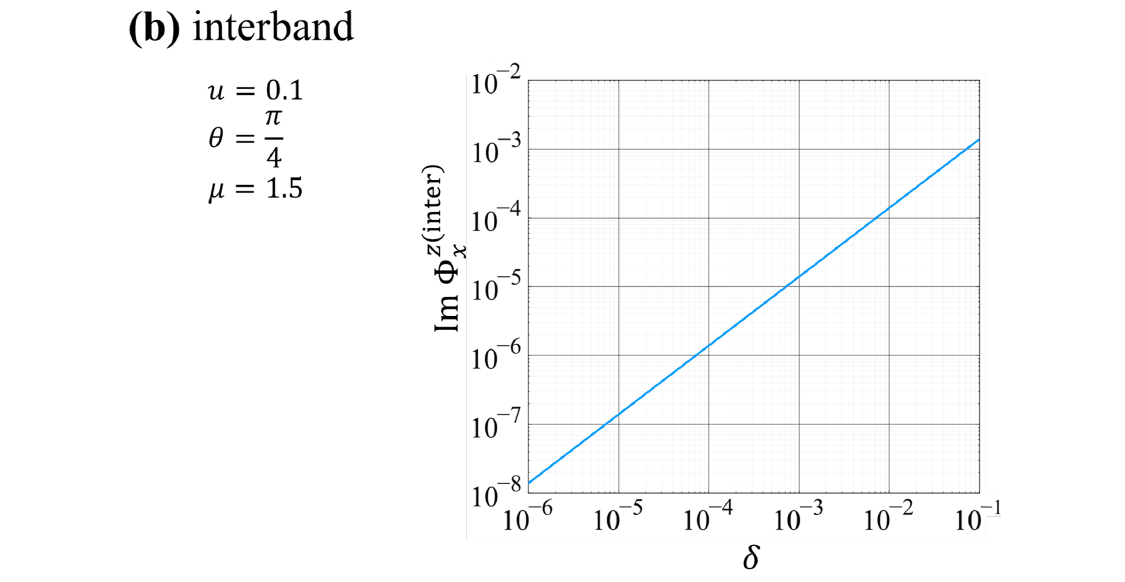} 
    \caption{(Color online) Broadening factor $\delta$ dependence of the spin current response for (a) the intraband and (b) interband components at $\mu=1.5$.}
    \label{fig:delta}
\end{figure}

The relative magnitude between the intraband and interband components depends on the broadening factor (relaxation time).
To demonstrate this, we plot the broadening factor dependence of ${\rm Im}[\Phi^{z{\rm (intra)}}_x]$ in Fig.~\ref{fig:delta}(a) and ${\rm Im}[\Phi^{z{\rm (inter)}}_x]$ in Fig.~\ref{fig:delta}(b) for fixed $\mu=1.5$; the other parameters are the same as those used in Fig.~\ref{fig:mu}. 
The data indicates the distinct $\delta$ dependence from each other; the intraband contribution is scaled as $\delta^{-1}$, whereas the interband contribution is scaled as $\delta^1$.
This indicates that the effect of the interband component is negligible compared to the intraband component in the clean limit ($\delta \to 0$), while it can be comparable when $\delta$ increases.
The different $\delta$ dependence for the intraband and interband processes arises from the factor $1/(\varepsilon_{m\bm{k}}-\varepsilon_{n\bm{k}}+i\delta)$ in Eq.~(\ref{eq:spin_current_response_2}), 
which is proportional to $\delta^{-1}$ for $\varepsilon_{m\bm{k}}=\varepsilon_{n\bm{k}}$ and approximately proportional to $\delta^{+1}$ for $\varepsilon_{m\bm{k}} \neq \varepsilon_{n\bm{k}}$ and $|\varepsilon_{m\bm{k}}-\varepsilon_{n\bm{k}}|> \delta$.

Reflecting that the noncollinear spin configuration is the essence of causing the antisymmetric spin splitting and spin current generation induced by elastic waves, their mean-field dependence is proportional to $h^2$~\cite{Hayami2020}.
This is also understood from the fact that the effective $z$-spin polarization is obtained by multiplying the $x$-spin and $y$-spin components in the mean field, as demonstrated in Ref.~\citen{Hayami2020}. 
Accordingly, even when considering the different domain for reversing $h \to -h$, the same sign of ${\rm Im} [\Phi^z_x] $ is obtained, meaning no cancellation of physical responses between the domain with $+h$ and $-h$. 
In this sense, this phenomenon is expected to be observed in experiments without the complication of preparing a single-domain state.

Furthermore, the emergence of ${\rm Im}[\Phi^{z}_x]$ and other components is intuitively understood as follows. 
Since the antisymmetric spin splitting contributes to ${\rm Im}[\Phi^{z}_x]$, as discussed above, its functional form determines nonzero components of the spin current. 
Indeed, the present antisymmetric spin splitting in the form of $k_y(k_y^2 - 3k_x^2)
\sigma_z$ is regarded as the coupling form given by 
\begin{align}
k_y(k_y^2 - 3k_x^2)\sigma_z &= -2k_x k_y \cdot k_x \sigma_z - (k_x^2 - k_y^2) \cdot k_y \sigma_z \nonumber \\
&=-k^2 \sin 2\theta \cdot k_x \sigma_z - k^2 \cos 2\theta  \cdot k_y \sigma_z,  
\end{align}
where we use $(k_x, k_y)=k (\cos \theta, \sin \theta)$ in the second line. 
From the symmetry correspondence, the first term is interpreted as the coupling between the spin current $J_x \sigma_{z}
$ and the $xy$-type strain, while the second term is interpreted as the coupling between the spin current $J_y\sigma_z$ and the $x^2-y^2$-type strain, where the former case corresponds to the results in Figs.~\ref{fig:mu} and \ref{fig:delta}. 
In the latter case, we can expect spin current generation in the $J_y\sigma_z$ component, which was confirmed by numerical calculations (not shown).

\begin{figure}[t]
    \centering
    \includegraphics[width=1.0\linewidth]{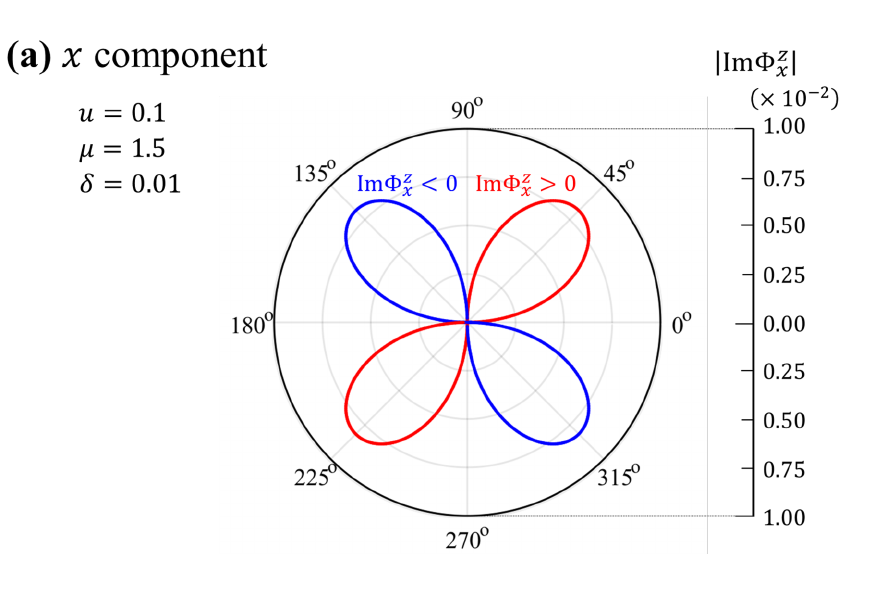} 
    \includegraphics[width=1.0\linewidth]{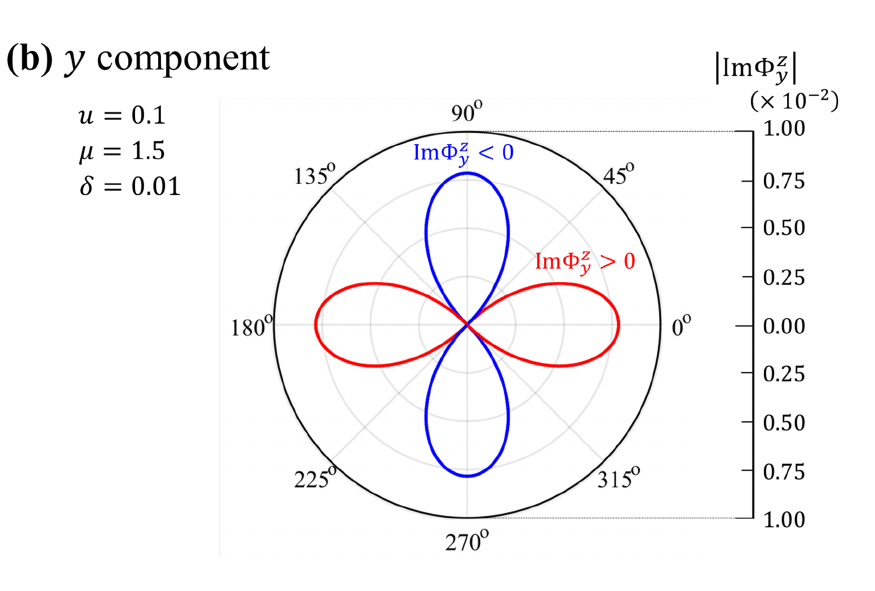} 
    \caption{(Color online) Angular dependence of the spin current response 
    induced by dynamical strain due to the application of a longitudinal wave 
    in the $f$-wave altermagnet, shown for (a) the $x$ component and (b) the $y$ component.}
    \label{fig:theta}
\end{figure}

From this analysis, we also expect the characteristic angle dependence of spin current generation; $J_x\sigma_z$ is generated by elastic waves with $\sin 2\theta$ dependence, while $J_y\sigma_z$ is generated with $\cos 2\theta$ dependence.
We numerically demonstrate such a feature by calculating the angle dependence of ${\rm Im}[\Phi^{z}_x]$ and ${\rm Im}[\Phi^{z}_y]$ for $u=0.1$, $\mu=1.5$, and $\delta=0.01$, as shown in Figs.~\ref{fig:theta}(a) and \ref{fig:theta}(b), respectively. 
The result indicates that the above intuitive picture is valid for the present mechanism of the spin current induced by elastic waves driven by the antisymmetric spin splitting in the band structure.

Finally, let us briefly discuss the comparison with the result for a nonmagnetic Rashba system, where a similar spin current generation by elastic waves was reported~\cite{Ogawa2023}.
As detailed in the Appendix, the present mechanism, driven by the magnetic phase transition, can yield a more efficient spin current response than the previous mechanism induced by the relativistic SOC. 
In addition, the present mechanism offers the advantage of controllable spin current generation, since the magnetic phase transition serves as a trigger for such responses. 
As the candidate materials for the two mechanisms differ, our present approach using magnetic materials provides an alternative route for inducing spin currents via elastic waves.

\section{Summary}
\label{sec:summary}
To summarize, we have theoretically investigated the mechanism of spin current generation induced by elastic waves driven by the magnetic-order phase transition
in an $f$-wave altermagnet that breaks the spatial inversion symmetry. 
Our analysis is based on the calculations for the tight-binding model with the three-sublattice noncollinear antiferromagnetic structure on the two-dimensional triangular lattice.
Taking into account the effect of lattice distortion induced by longitudinal elastic waves as hopping modulations, we evaluate the correlation function between the spin current and the strain within the linear response theory.
As a result, we find that the antisymmetric spin-split band structure triggered by the $f$-wave magnetic structure is significant in inducing the spin current generation by elastic waves even without the relativistic SOC. 
We also show that the intraband contribution becomes dominant in the clean limit, where the density of states becomes another important factor. 
Finally, we show the relationship between the antisymmetric spin splitting and the spin current induced by elastic waves from the symmetry viewpoint, which explains the characteristic direction dependence of the spin current response.

The present mechanism can be applied to other altermagnets breaking the spatial inversion symmetry, such as $p$-wave and $f$-wave altermagnets. 
The candidate material is the skyrmion-hosting material Gd$_3$Ru$_4$Al$_{12}$~\cite{hirschberger2019skyrmion, Hirschberger_10.1088/1367-2630/abdef9}, where the single-$Q$ spiral state corresponding to $p$-wave altermagnet appears at zero magnetic field~\cite{Yamada2025}.
Another candidate material is Ba$_3$MnNb$_2$O$_9$~\cite{Lee_PhysRevB.90.224402}, where it was revealed that the noncollinear spin configuration in a triangle unit exhibits the antisymmetric spin splitting even without the SOC~\cite{Hayami2020}. 
In addition, the materials with nearly 120$^{\circ}$ antiferromagnetic structures, such as CsFeCl$_3$~\cite{Hayashida_PhysRevB.97.140405} and PdCrO$_2$~\cite{mekata1995magnetic, ghannadzadeh2017simultaneous}, are also candidate materials. 
These findings highlight altermagnets as a novel platform for mechanically induced spin current generation, offering a pathway to spintronic functionalities without relying on relativistic effects, which might be promising for future applications to spintronic devices.

\begin{acknowledgments}
We express our gratitude to S. Miwa and K. Kondou for their fruitful discussions.
This research was supported by JSPS KAKENHI Grants Numbers JP22H00101, JP22H01183, JP23H04869, JP23K03288, JP23K2082703, and by JST CREST (JPMJCR23O4) and JST FOREST (JPMJFR2366).
\end{acknowledgments}

\appendix
\section{Comparison with nonmagnetic Rashba Systems}
\label{Rashba}
In this appendix, we briefly show the behavior of the spin current induced by elastic waves in a nonmagnetic Rashba system, which has been reported in Ref.~\citen{Ogawa2023}, and compare their results with the present ones.
For that purpose, we introduce the tight-binding model with the Rashba-type antisymmetric SOC instead of the magnetic mean field, which is given by 
\begin{equation}
H = H_{\mathrm{hop}} + H_{\mathrm{R}},
\end{equation}
where $H_{\mathrm{hop}}$ is the same as that in Eq.~(\ref{eq:Hhop}), although we adopt the single-sublattice form owing to the absence of magnetic ordering.
$H_{\mathrm{R}}$ denotes the Rashba-type antisymmetric SOC, which originates from the relativistic SOC in the absence of spatial inversion symmetry under the polar-type lattice structure. 
The explicit form of $H_{\mathrm{R}}$ is given by
\begin{align}
H_{\mathrm{R}} = i a_{\rm R}
 \sum_{\bm{k}\, \sigma,\sigma'} c_{\bm{k}\sigma}^\dagger \left[ \sum_{\bm{R}} e^{i \bm{k} \cdot \bm{R}} \left( \hat{\bm{R}} \times\bm{\sigma}_{\sigma\sigma'} \right) \right] c_{\bm{k}\sigma'},
\end{align}
where $a_{\rm R}$ is the Rashba coupling constant; we set to $a_{\rm R} = 0.1$ in the following calculation.
The vector $\hat{\bm{R}}$ denotes the unit bond vectors connecting nearest neighbors.

\begin{figure}[t]
    \centering
    \includegraphics[width=1.0\linewidth]{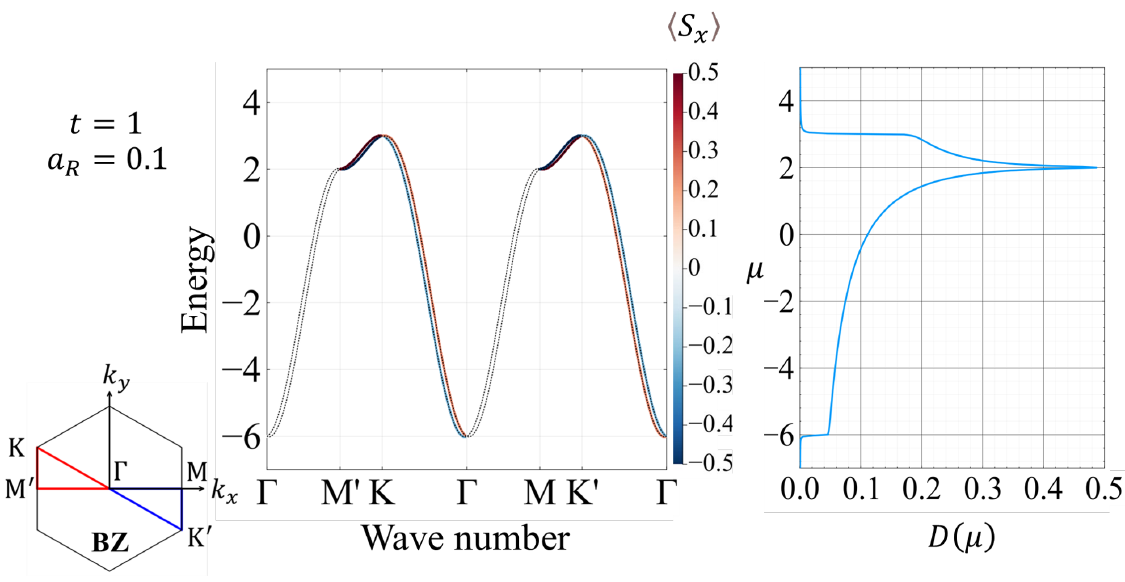} 
    \caption{(Color online) (Left panel) Antisymmetric spin-split band structure and (Right panel) the corresponding density of states in the Rashba system.
    The color in the left panel indicates the expectation value of the $x$-spin polarization.}
    \label{fig:band_r}
\end{figure}

The band structure for the Rashba model is presented in the left panel of Fig.~\ref{fig:band_r}, where the antisymmetric spin splitting in terms of the $x$-spin component is found in
the high-symmetry lines \({\rm M}\)--\({\rm K'}\)--\(\Gamma\) and \({\rm M'}\)--\({\rm K}\)--\(\Gamma\).
This indicates that the $x$-spin component of the antisymmetric spin splitting is represented by $k_y\sigma_x$. 
We also obtain the antisymmetric $y$-spin polarization for $k_x \neq 0$.
Thus, the total functional form of the antisymmetric spin splitting is given by $k_x \sigma_y - k_y \sigma_x$, which is consistent with the polar symmetry~\cite{rashba1960properties}. 
We also show the corresponding density of states in the right panel of Fig.~\ref{fig:band_r} for reference; the peak structure appears at $\mu=2$. 

\begin{figure}[t]
    \centering
    \includegraphics[width=1.0\linewidth]{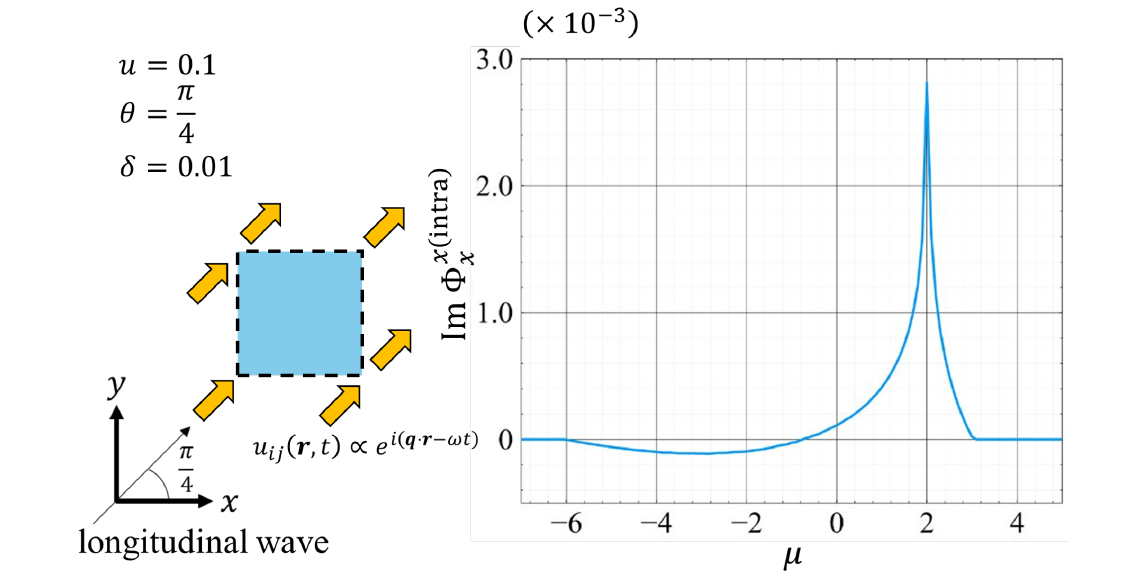} 
    \caption{(Color online) Chemical potential $\mu$ dependence of the intraband contribution to the spin current response induced by 
   longitudinal elastic waves in the Rashba system.}
    \label{fig:mu_r}
\end{figure}

The appearance of the antisymmetric spin splitting indicates the spin current generation induced by elastic waves, similarly to the case of $f$-wave altermagnets. 
We calculate the correlation function between the spin current and the strain in the nonmagnetic Rashba system under longitudinal elastic waves, as shown in Fig.~\ref{fig:mu_r}. 
In the calculations, we modulate the hopping amplitude in a similar manner in Eq.~(\ref{eq:modu_diff}) while keeping the Rashba term.
Among nonzero components in ${\rm Im}[\Phi^{\zeta}_i]$, we calculate ${\rm Im}[\Phi^{x}_x]$, which is induced by the Rashba-type antisymmetric spin splitting~\cite{Ogawa2023}. 
Similarly to the case of $f$-wave altermagnets, the intraband response is enhanced when the density of states becomes large. 
Meanwhile, no interband contribution was observed in the present system, which might be attributed to the single-sublattice degree of freedom. 
In addition, one finds that the magnitude of ${\rm Im}[\Phi^{x}_x]$ in the Rashba system is approximately one order of magnitude smaller than that in the $f$-wave altermagnetic system. 
This indicates that the $f$-wave altermagnet can be used as a generator of spin current induced by elastic waves, instead of the nonmagnetic Rashba system.

$^*$uchino@phys.sci.hokudai.ac.jp

$^\dagger$ogawa.y@phys.sci.hokudai.ac.jp

$^\ddagger$hayami@phys.sci.hokudai.ac.jp
\bibliography{71681.bib}
\bibliographystyle{jpsj}

\end{document}